\documentclass[useAMS,usenatbib]{mn2e}

\usepackage{graphics}
\usepackage{graphicx}
\usepackage{amsmath}
\usepackage{epsf}
\usepackage{gensymb}
\usepackage{rotating}
\usepackage{array}
\usepackage{subfigure}
\usepackage{longtable}
\usepackage{hyperref}
\usepackage{lscape}

\newcolumntype{C}[1]{ > {\centering\let\newline\\\arraybackslash\hspace{0pt}}m{#1}}
\newcolumntype{L}[1]{ > {\let\newline\\\arraybackslash\hspace{0pt}}m{#1}}

\def\reff@jnl#1{{\rm#1\/}}

\def\aj{\reff@jnl{AJ}} 
\def\araa{\reff@jnl{ARA\&A}} 
\def\apj{\reff@jnl{ApJ}} 
\def\apjl{\reff@jnl{ApJ}} 
\def\apjs{\reff@jnl{ApJS}} 
\def\ao{\reff@jnl{Appl.Optics}} 
\def\apss{\reff@jnl{Ap\&SS}} 
\def\aap{\reff@jnl{A\&A}} 
\def\aapr{\reff@jnl{A\&A~Rev.}} 
\def\aaps{\reff@jnl{A\&AS}} 
\def\azh{\reff@jnl{AZh}} 
\def\baas{\reff@jnl{BAAS}} 
\def\jrasc{\reff@jnl{JRASC}} 
\def\memras{\reff@jnl{MmRAS}} 
\def\mnras{\reff@jnl{MNRAS}} 
\def\pra{\reff@jnl{Phys.Rev.A}} 
\def\prb{\reff@jnl{Phys.Rev.B}} 
\def\prc{\reff@jnl{Phys.Rev.C}} 
\def\prd{\reff@jnl{Phys.Rev.D}} 
\def\prl{\reff@jnl{Phys.Rev.Lett}} 
\def\pasp{\reff@jnl{PASP}} 
\def\pasj{\reff@jnl{PASJ}} 
\def\qjras{\reff@jnl{QJRAS}} 
\def\skytel{\reff@jnl{S\&T}} 
\def\solphys{\reff@jnl{Solar~Phys.}} 
\def\sovast{\reff@jnl{Soviet~Ast.}} 
\def\ssr{\reff@jnl{Space~Sci.Rev.}} 
\def\zap{\reff@jnl{ZAp}} 
\def\nat{\reff@jnl{Nature}} 
\def\icarus{\reff@jnl{Icarus}} 
\def\acta{\reff@jnl{Acta Astron.}} 

\hypersetup{colorlinks=true,linkcolor=blue,citecolor=blue,filecolor=blue,urlcolor=blue}

\title[The detectability of habitable exomoons with {\it Kepler}]{The detectability of habitable exomoons with {\it Kepler}}
\author[S. Awiphan and E. Kerins]{S. Awiphan\thanks{E-mail: supachai.awiphan@students.manchester.ac.uk} and E. Kerins\\
Jodrell Bank Centre for Astrophysics, School of Physics and Astronomy, University of Manchester, Oxford Road, Manchester M13 9PL, UK}
\begin{document}

\date{Accepted 2013 April 10. Received 2013 April 9; in original form 2012 November 15}

\pagerange{\pageref{firstpage}--\pageref{lastpage}} \pubyear{2002}

\maketitle

\label{firstpage}

\begin{abstract}
In this paper, the detectability of habitable exomoons orbiting around giant planets in M-dwarf systems using Transit Timing Variations (TTVs) and Transit Timing Durations (TDVs) with {\it Kepler}-class photometry is investigated. Light curves of systems with various configurations were simulated around M-dwarf hosts of mass 0.5 $M_{\odot}$ and radius 0.55 $R_{\odot}$. Jupiter-like giant planets which offer the best potential for hosting habitable exomoons were considered with rocky super-Earth-mass moons. The detectability is measured by using the phase-correlation between TTV and TDV signals. Since the TDV signal is typically weaker than the TTV signal, confirmation of an exomoon detection will depend on being able to detect a TDV signal. We find that exomoons around planets orbiting within the habitable zone of an M-dwarf host star can produce both detectable TTV and TDV signatures with {\it Kepler}-class photometry. While aliasing between the planet period and moon period may hinder exomoon detection, we also find some strong correlation signatures in our simulation (eg. correlation: $>$0.7) which would provide convincing exomoon signatures. With the addition of red noise stellar variability, correlations generally weaken. However simulated examples with planet masses less than around 25 $M_{\oplus}$, moons of mass 8-10 $M_{\oplus}$ and specific values of planet and moon periods still yield detectable correlation in 25-50\% of cases. Our simulation indicates that {\it Kepler} provides one of the best available opportunities for exomoon detection.
\end{abstract}


\section{Introduction}

Over the last decade, the search for and study of exoplanets is one of the most dynamic research fields of modern astronomy. As of March 2013, more than 850 planets have been confirmed from ground-based observation\footnote{See \href{http://exoplanet.eu/}{\texttt{http://exoplanet.eu/}}} and more than 2,300 planet candidates have been discovered by the {\it Kepler} mission \citep{bat012}. As the number of detected exoplanets continues to grow, the potential for detecting satellites orbiting them has become of increasing interest in recent years \citep{sar999,han002,kip009a,kip009b,kip009kep}. The presence of exomoons may improve the probability of the existence of life on their host planet and the moons themselves also have potential to host life \citep{las993}. Moreover, the detection of exomoons would improve our understanding of planetary formation and evolution \citep{wil997}. 

Several methods have been developed to detect exomoons; including the transit \citep{sim007}, microlensing \citep{han002,lie010}, plusar-timing \citep{lew008}, Rossiter-McLaughlin effect \citep{sim010} and scatter-peak \citep{sim012}. An extension of the transit method involving transit timing appears to offer the best potential to detect habitable exomoons in the near future \citep{kip011b}. The presence of an exomoon induces two main variations on to the host planet; Transit Timing Variation (TTV) \citep{sar999} and Transit Duration Variation (TDV) \citep{kip009a,kip009b}. TTV and TDV signals are predicted to be $\pi$/2 out of phase for an edge-on circular orbit system, creating a unique exomoon signature. In order to claim the presence of an exomoon with the timing technique, both TTV and TDV signal must be detected.

Ideal host stars for habitable exomoon detection are M-type stars, due to the large amplitude of transits generated \citep{cha009,bea010tcs,vog010,man012} and the small distance of their habitable zone, which increases the transit probability and the number of transit events per observation time \citep{gai007,kal010}. Several studies (eg. \cite{bar002,dom006}) explore the stability of orbits around gas giants. In 2002, \citeauthor{bar002} proposed that Earth-mass moons of Jupiter-mass planets around stars of mass greater than 0.4 $M_{\odot}$ are dynamically stable for billions of years, long enough to sustain life.

At present, the best instrument up to this challenge is {\it Kepler} which launched in 2009. {\it Kepler} is designed to detect small transiting exoplanets with its highly sensitive photometric camera. It is monitoring $\sim$150,000 stars in Cygnus, including more than 3,000 M-dwarf stars \citep{bat010}. Many studies in exomoon detection with {\it Kepler} have been done (eg. \cite{sza006,kip012,sim012}), including a study of \cite{kip009kep} using TTV and TDV techniques that found that exomoons around gas giants planets in the habitable zone of M-dwarf stars may be detectable with the {\it Kepler} Mission. The Hunt for Exomoons with Kepler (HEK) project aims to search {\it Kepler} data for evidence of exomoons \citep{kip012}.

In this paper, we assess the detectability of habitable exomoons using the phase-correlation between TTV and TDV signals from simulation of {\it Kepler}-class photometry. The structure of the paper is as follows. In Section 2, a background of habitable exomoons with their timing effects is provided. In Section 3, the target selection is described. The generating of light curve of an exoplanet with an exomoon with {\it Kepler} is described in Section 4. The last part of this Section is dedicated to analysing the detectability of habitable exomoons using a detailed numerical simulation. In Section 5, simulation results are presented and analysed. Finally, the methods and results are summarised in Section 6.

\section{Habitable exomoon detection}
\label{sec:Exomoon}

\subsection{Habitable exomoons} 

An exomoon is a natural satellite of an exoplanet. None have been discovered to date. However, if our Solar System is typical, then exomoons must be common.

In 1997, \citeauthor{chy997} and \citeauthor{wil997} proposed the possibility of habitable exomoons. They proposed that moons orbiting around giant gas planets' Hill sphere could host life. Although moons orbiting giant planets at 1 AU from a solar analogue would become tidally locked within a few billion years after they form, their orbital period on timescales of a few days to a few months could cause temperature fluctuations on them \citep{chy997,wil997}. \cite{kip009kep} suggested that the habitable zone of exomoons can be defined simply as the distance where planets receive the same energy as the Earth,  

\begin{equation}
{a_{HZ}}=\sqrt{\frac{L_{*}}{L_{\odot}}} \ \mathrm{AU}. \ , 
\end{equation}
where $L_{*}$ is the host star luminosity. However, stellar evolution can alter the boundaries of the  habitable zone. Due to stellar evolution, the host star becomes brighter and hotter which shifts the habitable zone outward \citep{sel007}. For high-mass stars, their habitable zones are much further and broader which could be the best candidates for finding habitable planets or moons. However, emission in the far ultraviolet (FUV) band which is potentially damaging to life is observed in high-mass stars. While low-mass stars have the longest lifetimes, in their early age they have strong magnetic activity. Therefore, if planets or moons around the dwarfs can hold on to their atmosphere in the early age and are not tidally locked, they could be habitable \citep{jos997}.

Habitable moons need to be large enough to retain water and an atmosphere. Although moons formed from the proto-planetary disk are unlikely to be greater than 0.02\% the mass of the host planet \citep{can006}, moons formed from captures (Triton; \cite{agn006}) or impacts (the Moon; \cite{tay992}) have no limit on their mass. The results of \cite{wei010} show that there are some planets which could retain massive moons (Earth-size moon), including detected exoplanets such as CoRoT-3b and CoRoT-9b. 

Another major factor that should be taken into account is orbital stability. Because of three-body instability, moons can be lost from their host planet if the distance between the planet and moon is too large. To be retained by the planet, the moon must have an orbit that lies within the Hill sphere, $R_{H}$ \citep{bar002}.

\begin{equation}
R_{H}=a_{p}{(\frac{M_{p}}{3M_{*}})}^{1/3} \ ,
\end{equation}
where $M_{*}$ is mass of star, $M_{p}$ is mass of planet and $a_{p}$ is star-planet semi-major axis. In this project, the satellites within the Hill sphere of the habitable zone planets are considered, due to their habitability.

In the remainder of this paper, we use the terms "moon" and "exomoon" to refer to the less massive companion of a planet-satellite system even where the companion may itself be of planetary mass. 

\subsection{Transit Timing Variations (TTV)}
\label{sec:TTV}

In order to detect exomoons, several methods have been developed. The TTV and TDV techniques are focused upon in this work since they could detect small exomoons \citep{sar999,kip009a,kip009b}. The concept of the TTV technique is that the presence of a third body such as exomoon in the system causes a change in planetary orbit. The time between transits varies because the transiting planet and the moon exchange energy and angular momentum. This gravitational interaction perturbs the orbit of the transiting planet and causes a short-period oscillation of the semi-major axes and eccentricities. The signal depends on the mass, separation and orbital parameters of the planet and the moon (See \cite{kip011b}). For edge-on circular orbits, the TTV signal and RMS amplitude of the signal can be written as, 

\begin{equation}\label{eq:TTVC}
TTV= \left [ \frac{a_{m}M_{m}P_{p}}{2\pi a_{p}M_{p}} \right ]\cos (f_{m}) \ ,
\end{equation}
and

\begin{equation}\label{eq:TTVRMSC}
\delta_{TTV}=\frac{a_{m}M_{m}P_{p}}{a_{p}M_{p}\sqrt{2\pi}} \ ,
\end{equation}
where $M_{m}$ is the moon's mass and $a_{m}$ is the semi-major axis of the moon around the planet-moon barycentre, $P_{p}$ is the orbital period of the planet and $f_{m}$ is the true anomaly of the moon.
 
Unfortunately, TTV can also be induced by a multitude of phenomena, including general relativistic precession rate of periastra \citep{jor008}, stellar proper motion \citep{raf009} and parallax \citep{sch007} effects. Therefore, a TTV signal by itself cannot confirm the presence of an exomoon.

\subsection{Transit Duration Variation (TDV)}
\label{sec:TDV}

TDV is the periodic change in the transit duration over many measurements caused by the apparent velocity of the planet which increases and decreases due to the planet-moon interaction. \citeauthor{kip009a} showed that exomoons should induce not only the TTV effect but also the TDV effect on their host planets.

For the systems with non-coplanar orbits, the TDV effect can be separated into two main constituents, a velocity (V) component and a transit impact parameter (TIP) component \citep{kip009b}. The V-component is caused by the variation in velocity of the planet due to the moon's gravity. The TIP-component is affected by the planet moving between high and low host-star impact parameters. \cite{kip011b} formulated that the total TDV signal as a linear combination between TDV-V signal and TDV-TIP signals (See \cite{kip011b}).

In a system with 90 degree orbital inclination and a circular orbit, the TDV TIP-component will be zero. Therefore, the TDV signal of edge-on circular orbit is,

\begin{equation}\label{eq:TDVC}
TDV= \bar \tau\left [ \frac{a_{m}M_{m}P_{p}}{a_{p}M_{p}P_{m}}\right ]\sin (f_{m}) \ ,
\end{equation}
and the RMS amplitude of the TDV signal is,

\begin{equation}\label{eq:TDVRMSC}
\delta_{TDV}=\bar \tau \frac{a_{m}M_{m}P_{p}}{a_{p}M_{p}P_{m}\sqrt{2\pi}} \ ,
\end{equation}
where $P_{m}$ is the orbital period of the moon and $\bar \tau$ is transit duration. The TDV technique cannot detect habitable exomoons alone because the TDV signals are relatively weak compared with the TTV signals \citep{por011} and can also be induced by parallax effects \citep{sch007}. However, combining TDV and TTV signals can confirm the presence of an exomoon, because the signals have a $\frac{\pi}{2}$-phase difference that provides a unique exomoon signature. The orbital separation and mass of exomoons can also be obtained.

\section{Modelling habitable exomoons}

\subsection{Properties of the host star}

In this analysis, M-dwarf stars are selected to be the exoplanet host stars. Very cool (late K and early M type) dwarf stars have become popular targets of planet searches, because the amplitudes of the transits generated by planets in M-dwarfs are larger than those generated by hotter stars \citep{cha009, bea010tcs, vog010, man012} and the small distance of their habitable zone increases the transit probability of habitable planets as well as the transit frequency per observation time \citep{kal010}. \cite{sas012} also suggested that the semimajor axis of the host planet for the most detectable exomoons around an M-dwarf star is 0.2-0.4 AU. Therefore, the most detectable exomoons in M-dwarf systems can orbit within the habitable zone. However, the host with mass less than 0.2 $M_{\oplus}$ cannot host a habitable exomoon \citep{hel012}. 

{\it Kepler} monitored the Cygnus region along the Orion arm centred where there are about 0.5 million stars brighter than 16$^{th}$ magnitude ({\it Kepler} passband) within its FOV. However, only $10^5$ stars with magnitude less than 16 are expected to be exoplanet hosts. In 2010 the {\it Kepler} mission announced 150,000 highest priority target stars, but only 2\% of these target stars have effective temperature less than 3500 K \citep{bat010}, whereas $>$70\% of all stars within 20 pc are M-dwarfs \citep{hen994,cha003,rei004}. However, in 2011, the team released additional exoplanet data, including 997 planet-candidate host stars in which 74 ($>$5\%) have effective temperature less than 4400 K in the {\it Kepler} Input Catalog \citep{bat010, bor011, bro011}.

For our simulation of TTV and TDV signals, we assume that the host is an M-dwarf star with mass 0.5 $M_{\odot}$ and radius 0.55 $R_{\odot}$. Their effective temperature, microturbulent velocity and $\log g$ are set to be 3500 K, 1 km.s$^{-1}$ and 4.5, respectively, as applicable to solar-metallicity M-dwarf \citep{bea006,one012}. In order to calculate the limb darkening coefficient, solar-metallicity is assumed and a quadratic limb-darkening model is used. The values of limb-darkening coefficients for the transmission curves of {\it Kepler} are obtained from \cite{cla011}. For our M-dwarf targets, the value of the coefficients $\gamma_{1}$ and $\gamma_{2}$ are 0.4042 and 0.3268, respectively\footnote[2]{See \href{http://cdsarc.u-strasbg.fr/viz-bin/qcat?J/A+A/529/A75}{\texttt{http://cdsarc.u-strasbg.fr/viz-bin/qcat?J/A+A/529/A75}}}.

\subsection{Properties of the host planet}

Jupiter-like giant planets offer the best potential for detecting habitable exomoons \citep{kip009kep}. In order to investigate habitable exoplanets and exomoons, the planet-star separation is set to be inside the habitable zone, starting at a separation of 0.10 AU and increasing in logarithmically to 0.66 AU. This range includes the semimajor axis of M-dwarf planets which  \cite{sas012} argue on stability ground may be among first detectable exomoon systems.

We simulate giant planets with masses ranging logarithmically from 15 to 150 $M_{\oplus}$. \cite{for007} found that the radius of giant planets depends on their overall mass, core mass and separation. For giant planets of age 4.5 Gyr, their radius falls between 1.0 and 1.2 $R_{J}$ (Jupiter radius). Therefore, we adopt a planet radius of 1.2 $R_{J}$.
\subsection{Properties of the exomoon}

No exomoon has yet been discovered, therefore the properties of Earth-like planets are used for the habitable exomoon in this work. Rocky planets with logarithmic mass between 1 and 10 $M_{\oplus}$ are chosen. The radius of the moon is calculated from Fortney's model, using rock mass fraction equal to 0.66 (Earth-like planet):

\begin{equation}
R_{m}=1.00+0.65\log M_{m}+ 0.14(\log M_{m})^{2}
\label{MoonRadius}
\end{equation}
where $M_{m}$ and $R_{m}$ are the moon's mass and moon's radius in $M_{\oplus}$ and $R_{\oplus}$, respectively \citep{for007err, for007}. Only a moon within the planet's Hill sphere with an orbital period between 1.00 to 3.16 days is considered. Again, for simplicity, circular orbits are assumed.

\section{Generating and analysing light curves}

\subsection{{\it Kepler} transiting light curve generation}

In order to generate transit light curves of a planet with a moon, the algorithms of \cite{kip011} are used. The {\it Kepler} mission is designed to monitor $\sim$150,000 stars brighter than 16$^{th}$ magnitude (in the {\it Kepler} passband) with 20 parts per million photometric precision at 12$^{th}$ magnitude in 6.5 hours \citep{bat010, cal010k}. In order to meet this requirement, the estimated photon collection rate of {\it Kepler} is \citep{bor005, yee008, kip009kep}, 

\begin{equation}
\Gamma_{ph}=6.3\times 10^{8} \ \mathrm{h}^{-1} \ 10^{-0.4(m-12)} \ \mathrm{photons/hours}, 
\end{equation}
where $m$ is the apparent magnitude. However, {\it Kepler} photometry is also affected by shot noise, background flux and instrumental noise. Table~\ref{Keplernoise} summarise the properties we assume for {\it Kepler} photometry, including noise contributions we now discuss. 

\begin{figure*}
\begin{minipage}{17cm}
{\centering
\begin{tabular}{c}
\subfigure{\includegraphics[width=0.5\textwidth]{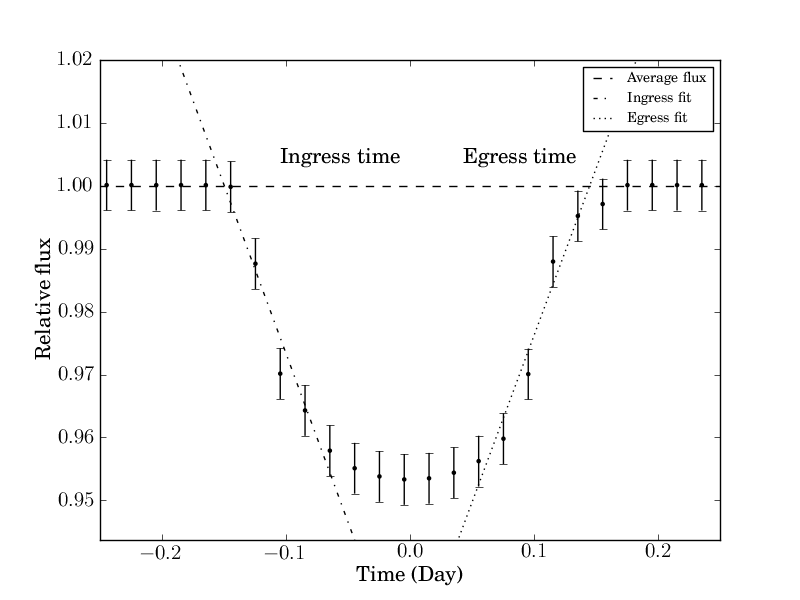}}
\subfigure{\includegraphics[width=0.5\textwidth]{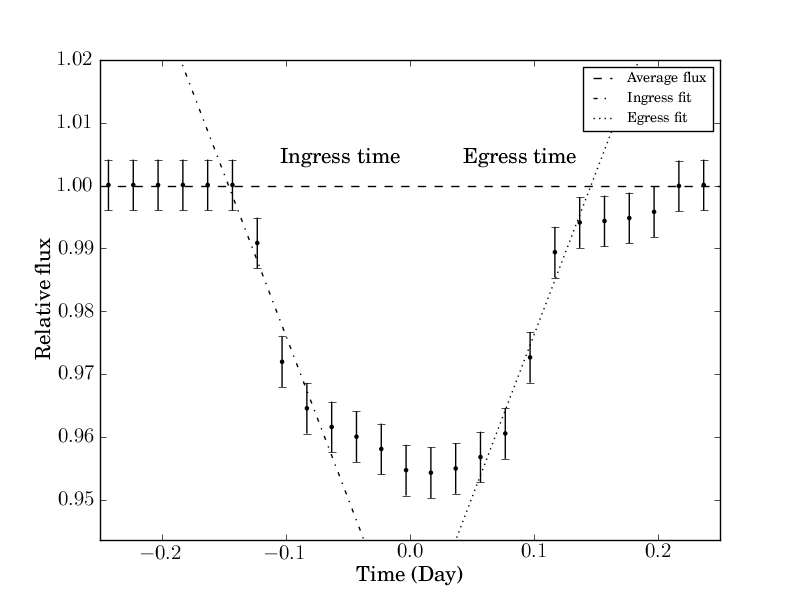}}
\end{tabular}}
\caption{Light curves of a 15 $M_{\oplus}$ habitable-zone planet with a 10 $M_{\oplus}$ moon for an M2 host star with planet period 89.35 days and moon period 2.24 days with differing moon phase. Error bars are shown at 1,000 times their true size. The fits show ingress time and egress time of transit, and average flux (median of flux data).}
\label{Fluxfit}
\end{minipage}
\end{figure*}

\subsubsection{Shot noise}

Shot noise or Poisson noise comes from the discrete nature of photons. At 12$^{th}$ magnitude, the largest noise component is the Poisson noise of the target \citep{cal010k} which we simulate.

\subsubsection{Background flux}

The background flux for {\it Kepler} comes from zodiacal light from the Solar System and diffuse starlight from background stars. In pre-launch prediction, the background flux is estimated at around 334 e$^{-}$sec$^{-1}$ or 22 magnitudes per square arcsecond \citep{cal010k}. However, in real observations, the background flux varies across detectors and with orientation of the telescope. We adopt the pre-launch background flux estimate to generate the light curves in this work.

\subsubsection{Instrumental noise}

There are two main components of instrumental noises for {\it Kepler}: read noise and dark current. From in-flight measurement, the read noise median value is 95 $e^{-}$read$^{-1}$ and dark current is 0.25 $e^{-}$pixel$^{-1}$s$^{-1}$ which is quite low compared to the photons collected from the targets \citep{cal010i}. We include it in our simulation, though its affect on our results is negligible.

\renewcommand{\thefootnote}{\fnsymbol{footnote}}

\begin{table}
\caption{{\it Kepler} photometry properties.}
\centering
{\footnotesize 
\begin{tabular}{C{4.5cm} C{1.5cm}}
\\
\hline\hline
\textbf{Parameter} & \textbf{Value} \\ [3pt]
\hline
Exposure time (s) & 6.02\footnote[1]{Obtain from \cite{kep009}} \\[1pt]
Plate scale (arcseconds/pixel) & 3.98\footnote[2]{Obtain from \cite{cal010k}} \\[1pt]
Background flux ($e^{-}$s$^{-1}$) & 334\footnotemark[2] \\[1pt]
Read Noise ($e^{-}$read$^{-1}$) & 95\footnote[3]{Obtain from \cite{cal010i}} \\[1pt]
Dark Current ($e^{-}$pixel$^{-1}$s$^{-1}$) & 0.25\footnotemark[3] \\[1pt]
\hline
\\
\multicolumn{2}{l}{$^*$ : \cite{kep009}} \\
\multicolumn{2}{l}{$^\ddagger$ : \cite{cal010i}} \\
\multicolumn{2}{l}{$^\dagger$ : \cite{cal010k}} \\
\end{tabular}
}
\label{Keplernoise}
\end{table}

\subsection{Measuring TTV-TDV signals}

In order to find the transit time of minimum and transit duration, the ingress and egress of our simulated light curves are fitted. The light curves are divided into phase bins using the input period which is assumed to be precisely determine from observational data. A running straight-line fit is made to three consecutive points of phased data. The fits with minimum and maximum slopes are chosen to define the ingress and egress of the transit, respectively. The intersection between the light curve median and the ingress and egress slopes are used to define the ingress ($t_{ing}$) and egress ($t_{egr}$) times, respectively (Figure~\ref{Fluxfit}). The time of minimum light ($t_{0}$) and the transit duration ($\bar{\tau}_{\textup{mean}}$) are defined as, $t_{0}=(t_{ing}+t_{egr})/2$, and $\bar{\tau}_{\textup{mean}}=t_{egr}-t_{ing}$, respectively.

Using the mid-transit time, a new ephemeris as a function of epoch is derived. The new ephemeris is determined by fitting a linear function to the mid-transit points.

\begin{equation}
T_{0}(n)=T_{0}(0)+nP \ , 
\end{equation}
where $n$ is epoch and $T_{0}$ is time of minimum light as a function of epoch. The residuals of the times of minimum and transit duration taken as the TTV and TDV signal of the system.

\subsection{TTV-TDV correlation testing}

\cite{hol005} showed that other planets could induce TTV signals on a transiting planet. Therefore, the TTV signal alone cannot distinguish between the effect of other planets and the effect of exomoon. In order to confirm an exomoon TDV signals must also be detected. 

From Section~\ref{sec:TDV}, TTV and TDV signals are sinusoidal functions and the TTV signal is 90 degrees out of phase with the TDV signal in coplanar systems. In the case of a circular planetary orbit and a co-aligned moon orbit, the TDV-TIP component exists. From Equation~\ref{eq:TTVC} and Equation~\ref{eq:TDVC}, the TTV signal, TDV signal and the relation between TTV and TDV are,

\begin{equation}\label{eq:TTVCo}
TTV= \left [ \frac{a_{m}M_{m}P_{p}}{2\pi a_{p}M_{p}} \right ]\cos (f_{m}) \ ,
\end{equation}

\begin{equation}\label{eq:TDVCo}
\setlength{\thinmuskip}{0mu}\setlength{\medmuskip}{0mu}\setlength{\thickmuskip}{0mu}{TDV= \bar \tau\left [ \left (\frac{a_{m}M_{m}P_{p}}{a_{p}M_{p}P_{m}}\right )+\left ( \frac{b_{p}}{1-b_{p}^{2}}\right )\left ( \frac{a_{m}M_{m}}{R_{*}M_{p}} \right ) \cos i_{p} \right ]\sin (f_{m})} \ ,
\end{equation}
and

\begin{align}
TDV^{2}&\setlength{\thinmuskip}{0mu}\setlength{\medmuskip}{0mu}\setlength{\thickmuskip}{0mu}{=- \left (\frac{2 \pi a_{p}\bar \tau}{P_{p}} \right )^{2} \left (\frac{P_{p}}{a_{p}P_{m}}+\frac{b_{p}}{1-b_{p}^{2}}\frac{\cos i_{p}}{R_{*}} \right )^{2} TTV^{2}} \nonumber\\
&\setlength{\thinmuskip}{0mu}\setlength{\medmuskip}{0mu}\setlength{\thickmuskip}{0mu}{+\left (\frac{a_{m}M_{m}\bar \tau}{M_{p}}\right )^{2} \left (\frac{P_{p}}{a_{p}P_{m}}+\frac{b_{p}}{1-b_{p}^{2}}\frac{\cos i_{p}}{R_{*}}\right )^{2}} \ ,
\end{align}
where $b_{p}$ is impact parameter of the planet from its host and $R_{*}$ is host star radius. However, the ratio between TDV-V and TDV-TIP is very large. For the systems in our simulation, the minimum ratio is 1,100. Therefore, the TDV-TIP is negligible and the relation between TTV and TDV signal can be written as,

\begin{equation}
TDV^{2}=-\left (\frac{2\pi \bar \tau}{P_{m}}\right )^{2} TTV^{2}+\bar \tau^{2} \left ( \frac{a_{m}M_{m}P_{p}}{a_{p}M_{p}P_{m}} \right )^{2} \ .
\end{equation}

Therefore, in theory, the plot between the square of the TTV signal and square of the TDV signal should show a perfect linear relationship with negative slope. However, there are other effects, such as star spots, instrument noise and sparsity of observation which could produce false positive TTV and TDV signatures.

In this simulation, the instrument noise and observing frequency both affect the TTV and TDV signals. Thus, the plot between $TTV^{2}$ and $TDV^{2}$ may not show a clear linear relationship. In order to check this relationship, the Pearson product-moment correlation coefficient was calculated to test the correlation between $TTV^{2}$ and $TDV^{2}$. The coefficient is

\begin{equation}
\chi=\frac{\sum_{i=1}^{n}(TTV^{2}_{i}-\overline{TTV^{2}})(TDV^{2}_{i}-\overline{TDV^{2}})}{\sqrt{\sum_{i=1}^{n}(TTV^{2}_{i}-\overline{TTV^{2}})^{2}}\sqrt{\sum_{i=1}^{n}(TDV^{2}_{i}-\overline{TDV^{2}})^{2}}} \ .
\end{equation}

\begin{figure*}
\begin{minipage}{17cm}
{\centering
\begin{tabular}{C{0.3cm} C{5.2cm} C{5.2cm} C{5.2cm}}
& \textbf{TTV} & \textbf{TDV} & \textbf{Correlation} \\
\end{tabular}
(a)
\begin{tabular}{c}
\subfigure{\includegraphics[width=5.5cm]{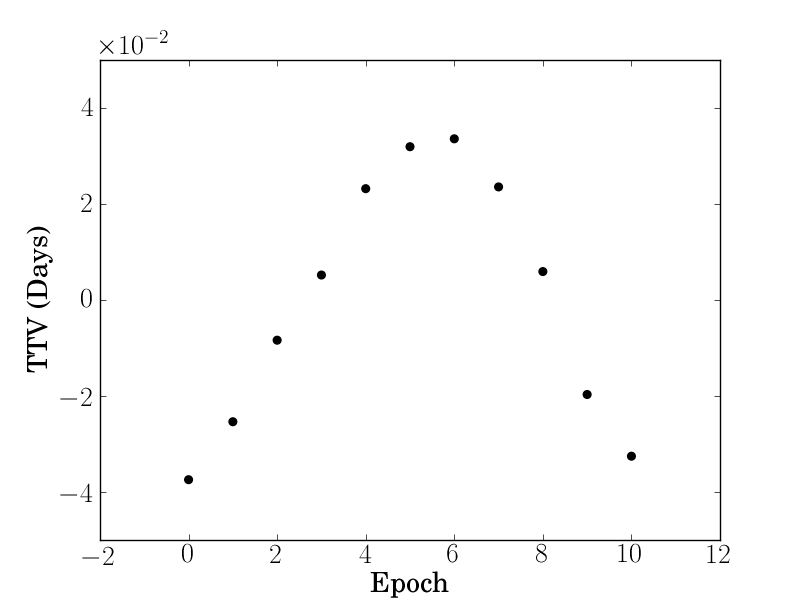}}
\subfigure{\includegraphics[width=5.5cm]{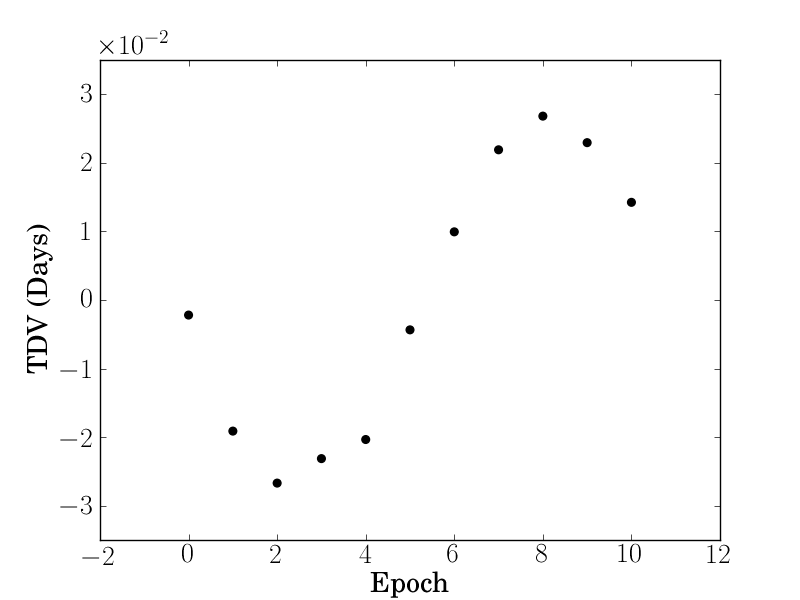}}
\subfigure{\includegraphics[width=5.5cm]{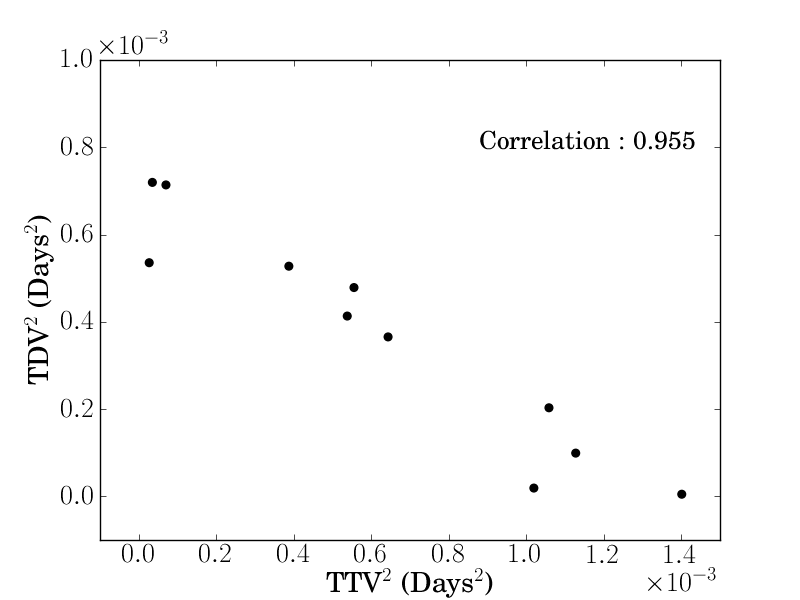}}
\end{tabular}
(b)
\begin{tabular}{c}
\subfigure{\includegraphics[width=5.5cm]{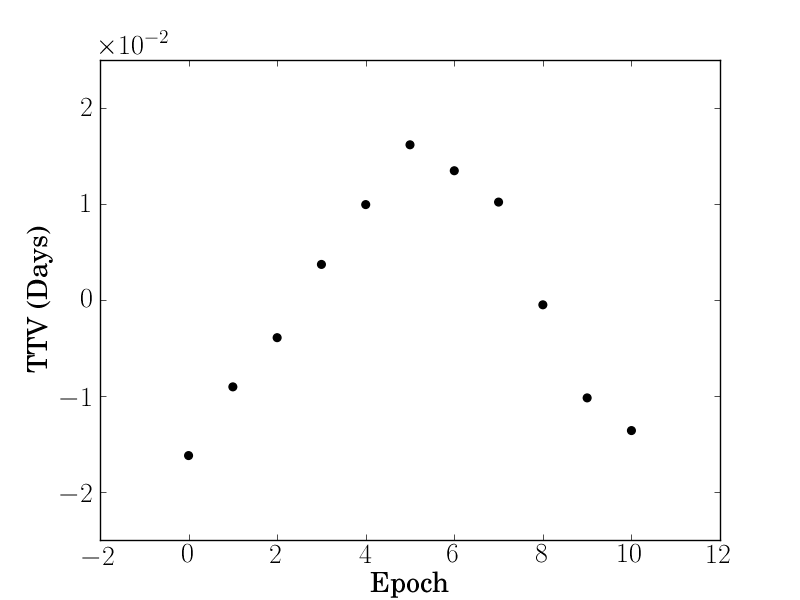}}
\subfigure{\includegraphics[width=5.5cm]{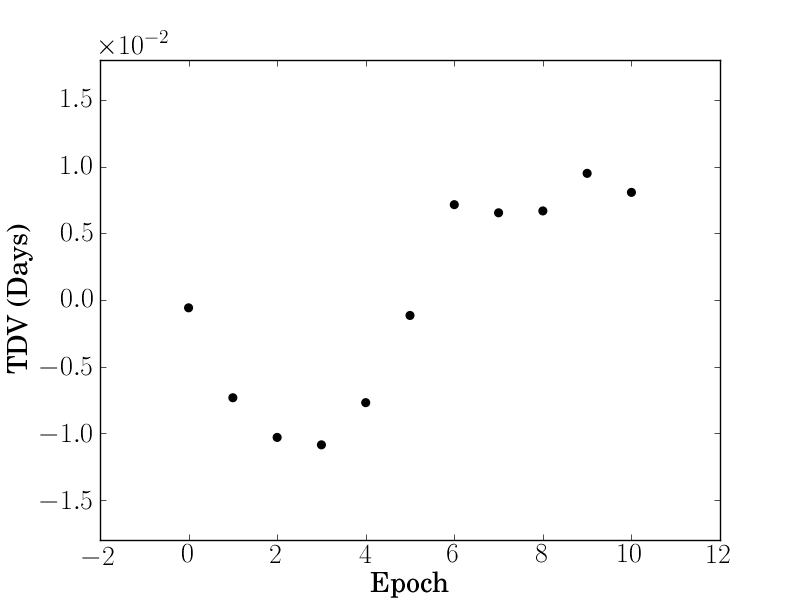}}
\subfigure{\includegraphics[width=5.5cm]{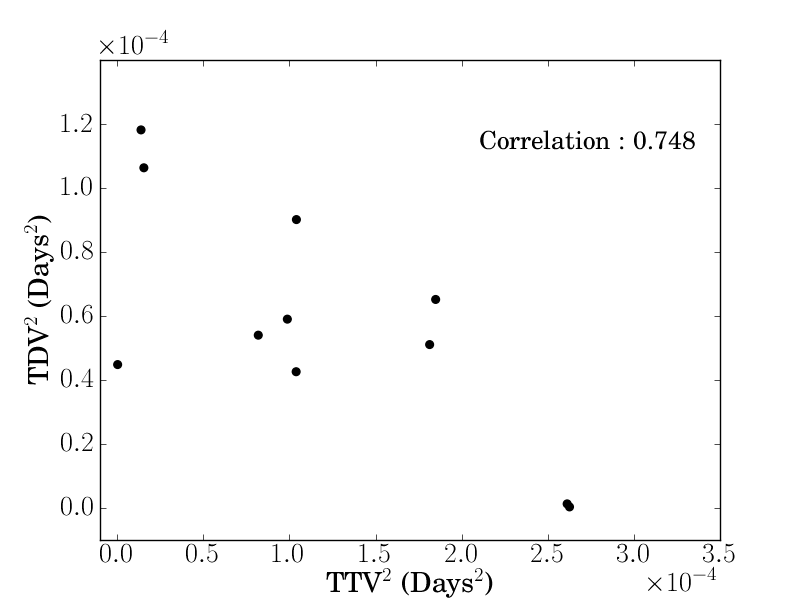}}
\end{tabular}
(c)
\begin{tabular}{c}
\subfigure{\includegraphics[width=5.5cm]{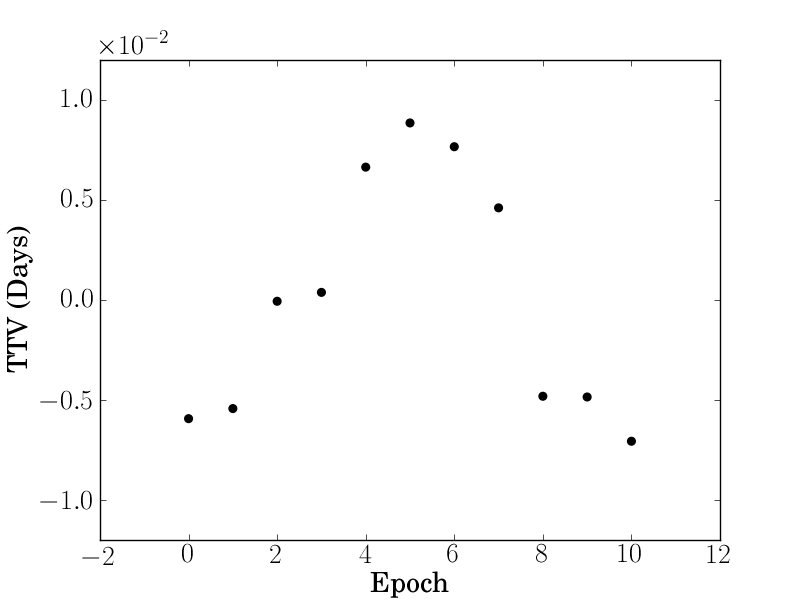}}
\subfigure{\includegraphics[width=5.5cm]{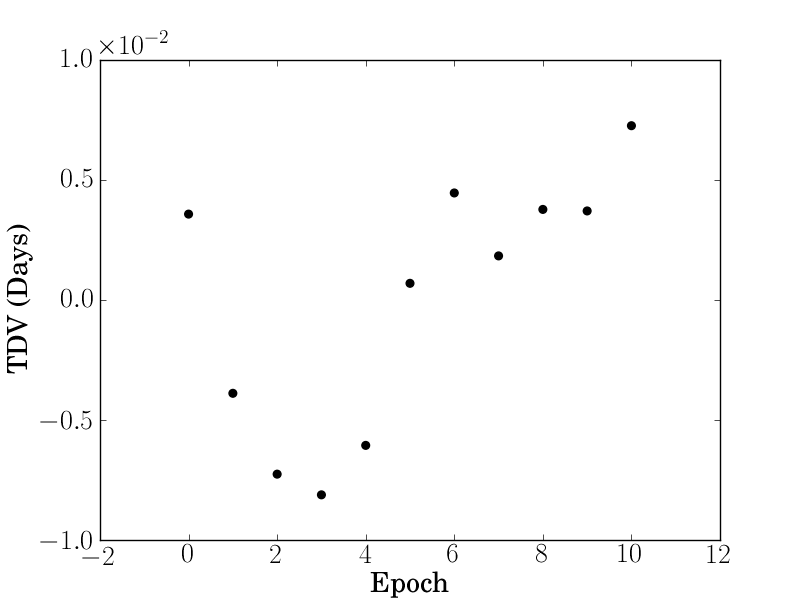}}
\subfigure{\includegraphics[width=5.5cm]{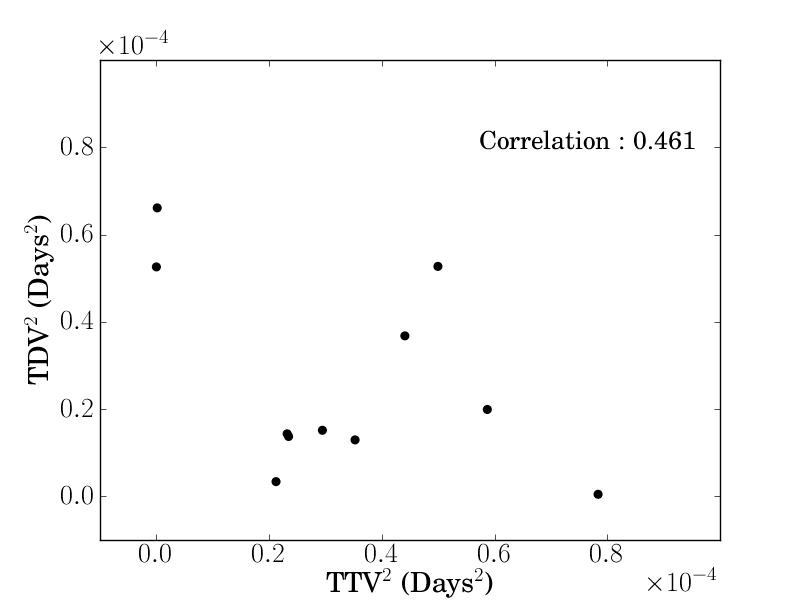}} 
\end{tabular}}
\caption{TTV signal (left), TDV signal (middle) and $TTV^2$ versus $TDV^2$ (right) for a 10.0 $M_{\oplus}$ exomoon of a planet orbiting around a 0.5 solar-mass M-dwarf star.  The with planet period is  89.35 days and moon period is 2.24 days. Panels are shown in three exoplanet with $\log(M_{p}/15M_{\oplus})$ equal to 0.0 (a), 0.4 (b), and 0.7 (c).}
\label{Cor}
\end{minipage}
\end{figure*}

A negative coefficient is produced by an inverse relationship between the two variables and a positive coefficient means there is a positive linear relationship. The positive slope of $TTV^{2}$ and $TDV^{2}$ plot means the TTV and TDV signal are not consistent with sinusoidal functions with a 90 degree phase difference. The TTV signal, TDV signal and $TTV^2$ versus $TDV^2$ of three sample systems are shown in Figure~\ref{Cor}.

\section{Detectability of habitable exomoons}

\subsection{Detectability of habitable exomoons}
\label{sec:DetectSec}

The light curves are generated with {\it Kepler} photometric noise. 146,410 light curves are simulated with 11 independent values of each of four variable input parameters: planet mass; planet separation; moon mass; and moon period, and $10$ random initial phases. The host stars are assumed to be M-dwarf stars of 12.5 magnitude in the {\it Kepler} passband. The cadence of this simulation is 50 data points per day (every 28.8 mins) which corresponds closely to {\it Kepler}'s long cadence mode (every 29.4 mins) \citep{gil010}. In order to simulate the current {\it Kepler} data, a 3-year simulation of a transiting giant extrasolar planet with a rocky extrasolar moon was run to find out the detectability of an exomoon in the M-dwarf habitable zone. The details of physical parameters of the systems are listed in Table~\ref{InputParameter}.

\begin{table}
\caption{Input parameters assumed for our exomoon simulations.}
\centering
{\footnotesize 
\begin{tabular}{L{5cm} C{2cm}}
\\
\hline\hline
\multicolumn{2}{c}{\textbf{Star parameters}} \\
\hline
Mass ($M_{\odot}$) & 0.5 \\[3pt]
Radius ($R_{\odot}$) & 0.55 \\[3pt]
Apparent magnitude ($K_{p}$) & 12.5 \\[3pt]
Quadratic limb-darkening coefficient 1 & 0.4042 \\[3pt]
Quadratic limb-darkening coefficient 2 & 0.3268 \\[3pt]
\hline
\multicolumn{2}{c}{\textbf{Planet parameters}} \\
\hline
Mass ($M_{\oplus}$) & 15.0-150.0 \\[3pt]
Radius ($R_{J}$) & 1.2 \\[3pt]
Separation (AU) & 0.10-0.66 \\[3pt]
Eccentricity & 0.0 \\[3pt]
Inclination (degrees) & 90.0 \\[3pt]
\hline
\multicolumn{2}{c}{\textbf{Moon parameters}} \\
\hline
Mass ($M_{\oplus}$) & 1.0-10.0 \\[3pt]
Radius ($R_{\oplus}$) & Equation~\ref{MoonRadius} \\[3pt]
Period (days) & 1.00-3.16 \\[3pt]
Eccentricity & 0.0 \\[3pt]
Inclination (degrees) & 90.0 \\[3pt]
\hline
\end{tabular}
}
\label{InputParameter}
\end{table}

The 4D-simulation is projected on to two-parameter planes in order to examine the relation between two variables. Since we are only interesting in negative correlations, we define the projected correlation as:
\begin{equation}
\chi _{\textrm{proj}}=\frac{1}{N}\sum _{i(\chi > 0)} ^{N} \chi _{i} \ , 
\label{chi}
\end{equation} 
where $N$ is the total number of 2-D simulations that are projected and $i(\chi >0)$ refers only to those simulations with negative correlation. The projected plots therefore represent averages over logarithmic parameter priors for negative correlation signals. In the left hand panels of Figure~\ref{MPMMkp}, the plot between planet mass and moon mass shows that a high-mass moon hosted by a low-mass planet is the most detectable of the systems considered. This result agrees with the moon period versus planet mass and moon mass plots (Figure~\ref{MPPMkp} and Figure~\ref{MMPMkp}). However, in these two plots, the changes in moon period do not affect the correlation. In Figure~\ref{PPPMkp}, the projection plot between separation of planet and period of moon also does not show any significant trends.

Figure~\ref{MPPPkp} shows the detectability coefficient between mass and separation of the planet. Planets with high separation have higher detectability than close-in planets of the same mass. This result correlates with the result of moon mass versus planet separation (Figure~\ref{MMPPkp}) which shows that, in systems of equal satellite mass, the outer planet hosts have larger correlation coefficients. These features may be produced by only a few transit events in high planet separation systems, because, at 0.6 AU separations, only three transit events are detected in the simulation. Therefore, we now check the reliability of the correlation.
 
The analysis of correlations is meaningful when the correlations are not dominated by noise. The variance of correlation is plotted in the right-hand panels of Figure~\ref{contourKP} in order to check the reliability of testing. From Figure~\ref{MPMMkp} to Figure~\ref{MMPPkp}, the variance plots show that the systems with a small number of transit events (long planet period systems) have higher variance. However, the value of the variance is still low compared to the correlation coefficient. 

While the magnitude of $\chi_\textrm{proj}$ in Equation~\ref{chi} is reduced by positive correlations that are included in $N$, we have checked that the basic features in the plots of Figure~\ref{contourKP} trace those obtained by ignoring positive correlations, albeit at a weaker level. Finally, the assumption that moons of outer planets should be easier to detect than moons of inner planets is confirmed by our simulations.

The theoretical lines of RMS amplitude of the TTV and TDV signals are shown to investigate the features of the contours. For Figure~\ref{MPMMkp} the high amplitude of TTV and TDV signals produces a high coefficient of detection with the same slope. Moreover, the features in Figure~\ref{MPPPkp} and Figure~\ref{MMPPkp} are also well-correlated with TDV RMS amplitude signals which can be explained by the relative weakness of TDV signals compared with TTV signals. In conclusion, the detectability of exomoons is dominated by the amplitude of TDV signals.

\subsection{Analysing the correlation structure}
\label{sec:AnalyseSec}

The structures of the correlation plots are explained by the magnitude of the TDV signal. However, in Figure~\ref{MPPPkp} and Figure~\ref{MMPPkp}, gaps are evident at planet semi-major axes of 0.4 and 0.5 AU. The variance plots show that there is no difference in variance across this region and that therefore the features in Figure~\ref{MPPPkp} and Figure~\ref{MMPPkp} are real. To investigate the structures, the 4-D plots are sliced into 2-D plots. The correlation plots of planet separation versus planet mass, and of planet separation versus moon mass are shown in Figure~\ref{MPPP} and Figure~\ref{MMPP}. Systems with high moon mass and low planet mass have a high value of correlation and the features in these contours correspond to the projected contours of Figure~\ref{contourKP}, including the gap structures.

In Figure~\ref{MPPP} and Figure~\ref{MMPP}, the maps with similar moon period show gap features at the same planet separation, but they shift with a different moon period. The ratios between moon period and planet which produce the gap are near-integer values and correspond to the cold spots in Figure~\ref{PPPMkp}. Therefore, the gap structures can be explained by a resonance between the planet and moon period which produces constant detected TTV and TDV signals. However, they also depend on the number of detected transit events. In short period systems which have a larger number of transits, the gap structures are more difficult to produce due to the larger range of detected planetary phases.

\subsection{M-dwarf variability}
\label{sec:RedNoiseSec}

\begin{figure}
\includegraphics[width=0.5\textwidth]{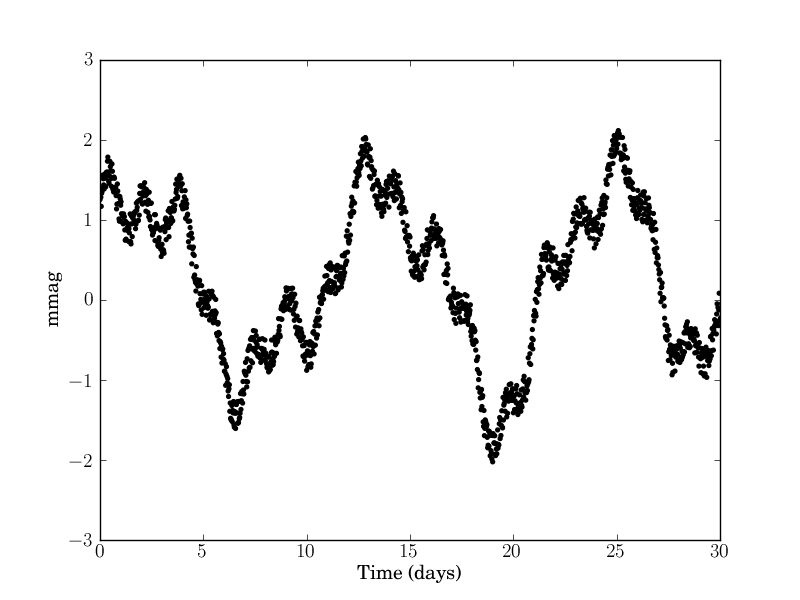}
\caption{Our simulated stellar red noise with a main noise component of 12 days period (See Section~\ref{sec:RedNoiseSec}).}
\label{VarLC}
\end{figure}

In the previous section, the light curves without intrinsic stellar variability were simulated. In reality, stars can vary in in brightness due to pulsation, rotation and activity (red noise). Around 40-70\% of M-dwarfs have variability with photometric dispersion ($\sigma_{m}$) $\sim$3-5 mmag, depending on their brightness. For stars with 12.5 magnitude in the {\it Kepler} passband, the variability fraction is nearly 1 \citep{cia011} and their noise tends to have long variation periods ($\geq$5 days) \citep{mcq012}. Therefore, in our simulation, the red noise with a 12 day period was added into the light curves in order to investigate the effect of stellar noise to the detectability. Their amplitude based on that found for M-dwarfs with {\it Kepler} magnitude between 12-14 \citep{cia011}. In order to simulate short term variability, five minor variations with linearly random periods between 0 and 12 days and amplitudes less than half of the amplitude of main variation were also added. An example of our simulated red noise is shown in Figure~\ref{VarLC}.

\begin{figure}
\includegraphics[width=0.5\textwidth]{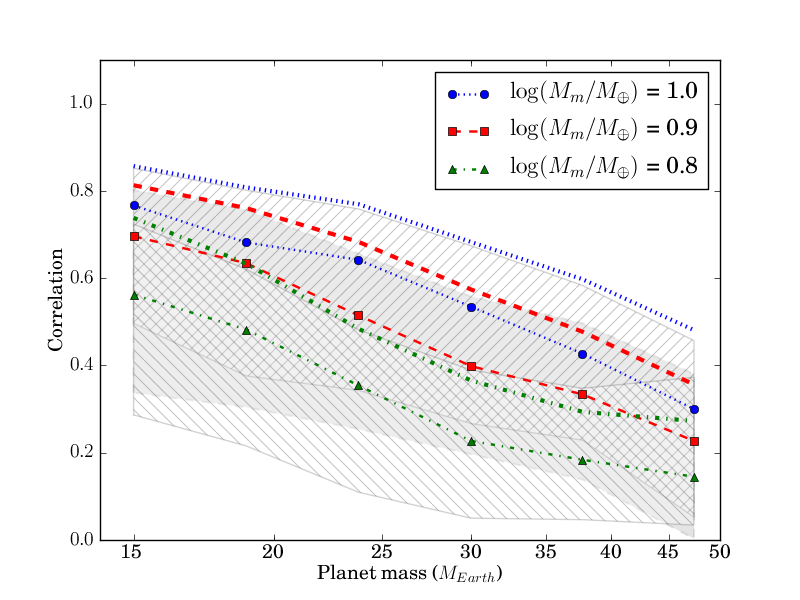}
\caption{Median correlation as a function of planet of mass $\log(M_{p}/15M_{\oplus})$ equal to 1.0 $M_{\oplus}$ (Blue circle with dot line), 0.9 $M_{\oplus}$ (Red square with dashed line) and 0.8 $M_{\oplus}$ (Green triangle with dashed-dot line) exomoon orbiting around a M-dwarf with planet period and moon period are 89.35 and 2.24 days, respectively. Thick lines show the median correlations of the systems without stellar variability. The forward diagonal hatch region, shaded area and backward diagonal hatch region represent the 25th to 75th percentile region of systems with 1.0, 0.9 and 0.8 $M_{\oplus}$ planets, respectively.}
\label{Var}
\end{figure}

Our test of the effects of red noise were based on following parameters. We simulated planets with $\log(M_{p}/15M_{\oplus})$ equal to 0.0, 0.1, 0.2, 0.3, 0.4 and 0.5 and tested moons with $\log(M_{m}/M_{\oplus})$ equal to 0.8, 0.9 and 1.0 $M_{\oplus}$. The planet and moon are assumed to have a period of around 89 and 2.2 days, respectively, that correspond to the peak in Figure~\ref{MPPP} and Figure~\ref{MMPP}. For each parameter combination, we used simulations with high TTV-TDV correlation ($>$0.7) to examine the effect of red noise on high-confidence detections. We added 500 different variations for each dispersion to each light curve. The result in Figure~\ref{Var} shows that the presence of intrinsic stellar variability of M-dwarfs might effect the exomoon detectability. The stellar variability reduces the exomoon detection correlation by 0.1. However, for our simulated systems with planet masses less than around 25 $M_{\oplus}$ with moon masses 8-10 $M_{\oplus}$, typically 25-50\% of them still have correlations high enough to confirmed exomoon detection.

\subsection{Improving the detectability with long-term observation}

Section~\ref{sec:DetectSec} and ~\ref{sec:AnalyseSec} are based on 1,000-day simulations of 50,000 data points (every 28.8 minutes), which corresponds to the integration time of the {\it Kepler} long-cadence strategy. In theory a longer time baseline should produce a better detectability coefficient, because of more transiting events. In 2012, {\it Kepler} has been approved for extension through to 2016 . Therefore, 5-years photometric data from {\it Kepler} targets should be obtained. We have performed a 1,600-day simulations to produce a sample of long-term observations. The moon period is set to be 2.51 days, because it is not in resonance with any planetary period considered in this work.

\begin{figure*}
\begin{minipage}{16.5cm}
{\centering
\begin{tabular}{c}
\subfigure{\includegraphics[width=16.5cm]{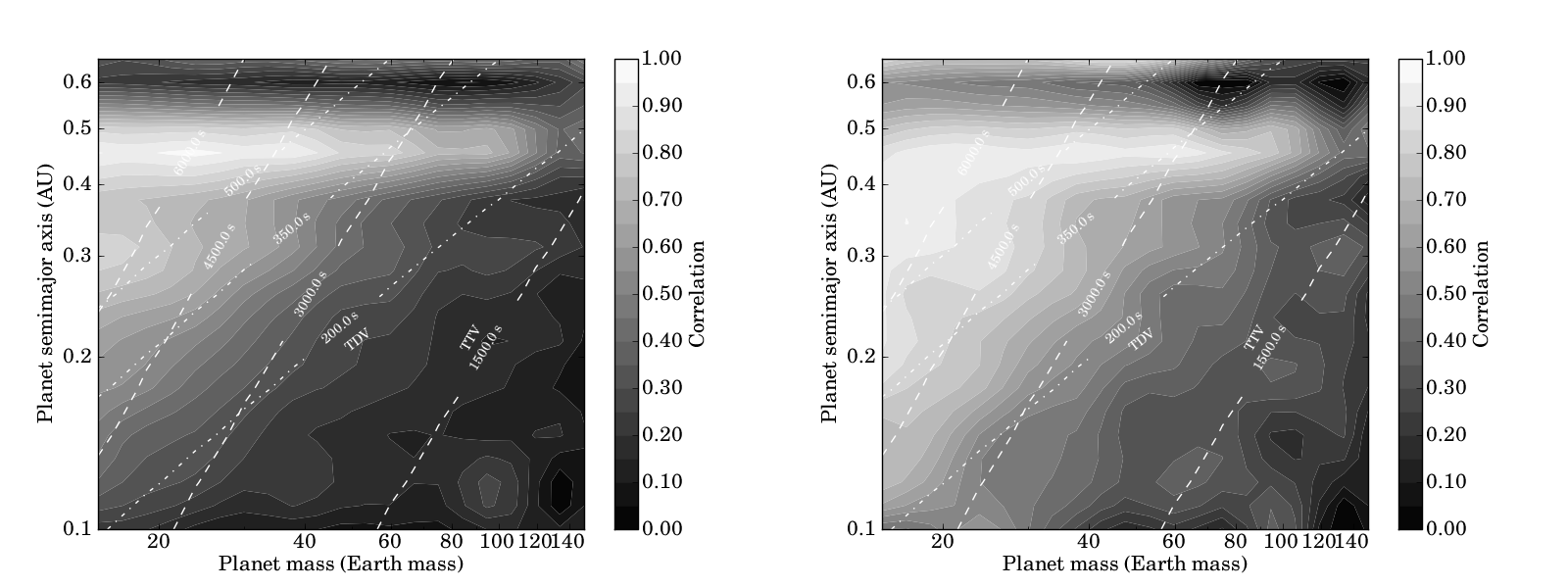} \label{MPPPkpL}} \\
(a) Planet mass versus planet semimajor axis \\
\subfigure{\includegraphics[width=16.5cm]{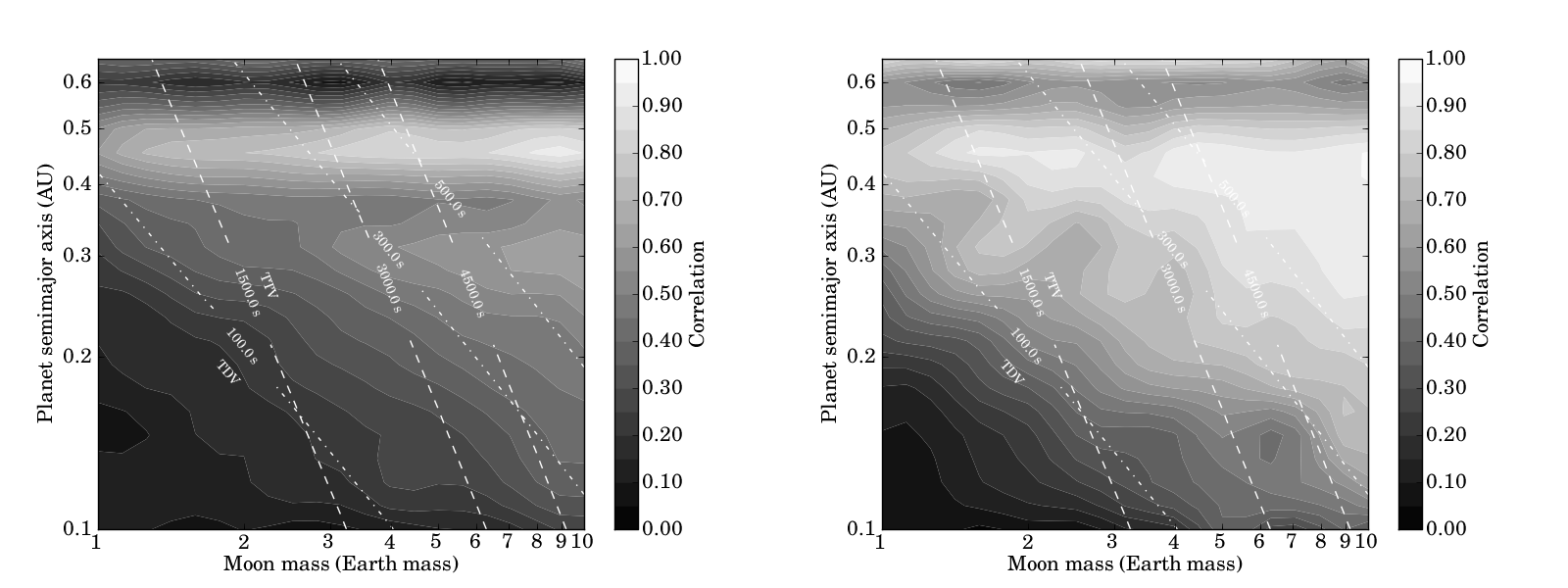} \label{MMPPkpL}} \\
(b) Moon mass versus planet semimajor axis \\
\end{tabular}}
\caption{Correlation contours between (a) planet mass and planet semimajor axis and (b) moon mass and planet semimajor axis for a 3-year simulation (left) and 5-year simulation (right). The moon period is set to be 2.51 days.}
\label{contourKPL}
\end{minipage}
\end{figure*}

The contours of two simulations show the same structures in Figure~\ref{contourKPL}, but the long-term simulation has a better detectability. From this result, the number of detected exomoons should increase with length of observation. The extension to the {\it Kepler} mission and possible additional space-based telescope missions (eg. PLATO \citep{rau012}) should allow the habitable exomoons to be detected or their abundance constrained.

\newpage

\begin{figure*}
\begin{minipage}{16.5cm}
{\centering
\begin{tabular}{c}
\subfigure{\includegraphics[width=16.5cm]{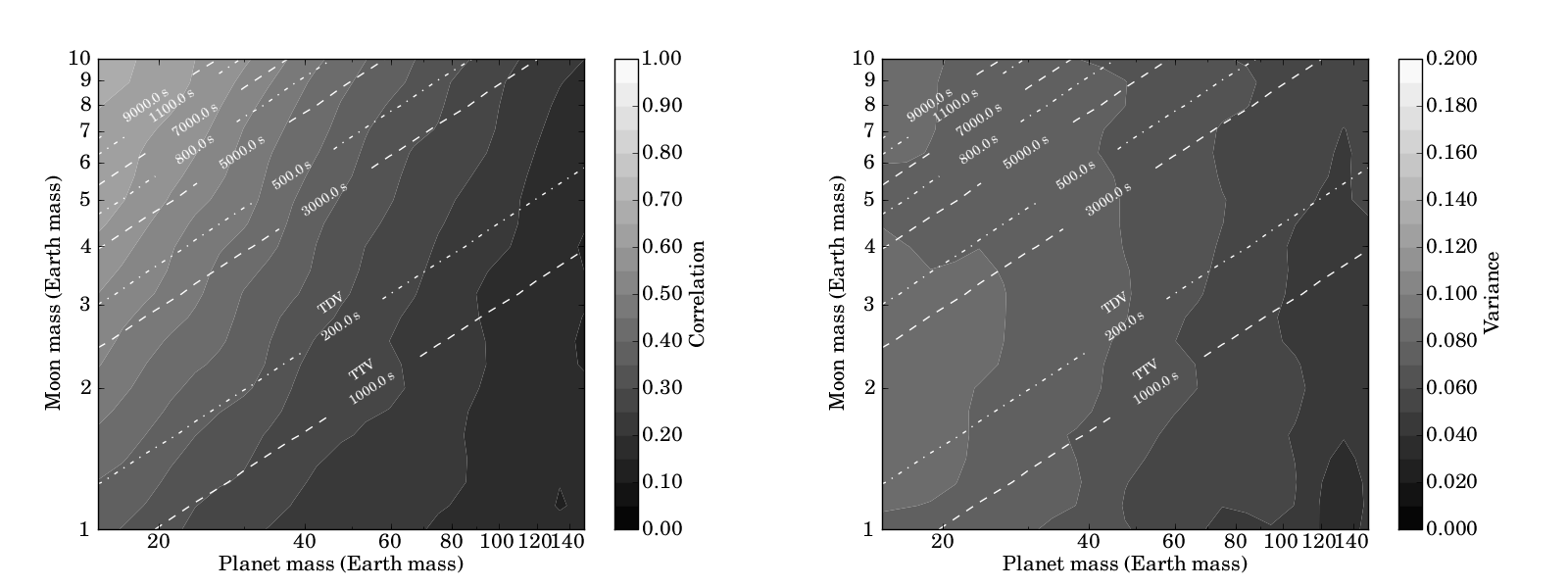} \label{MPMMkp}} \\
(a) Planet mass versus moon mass \\
\subfigure{\includegraphics[width=16.5cm]{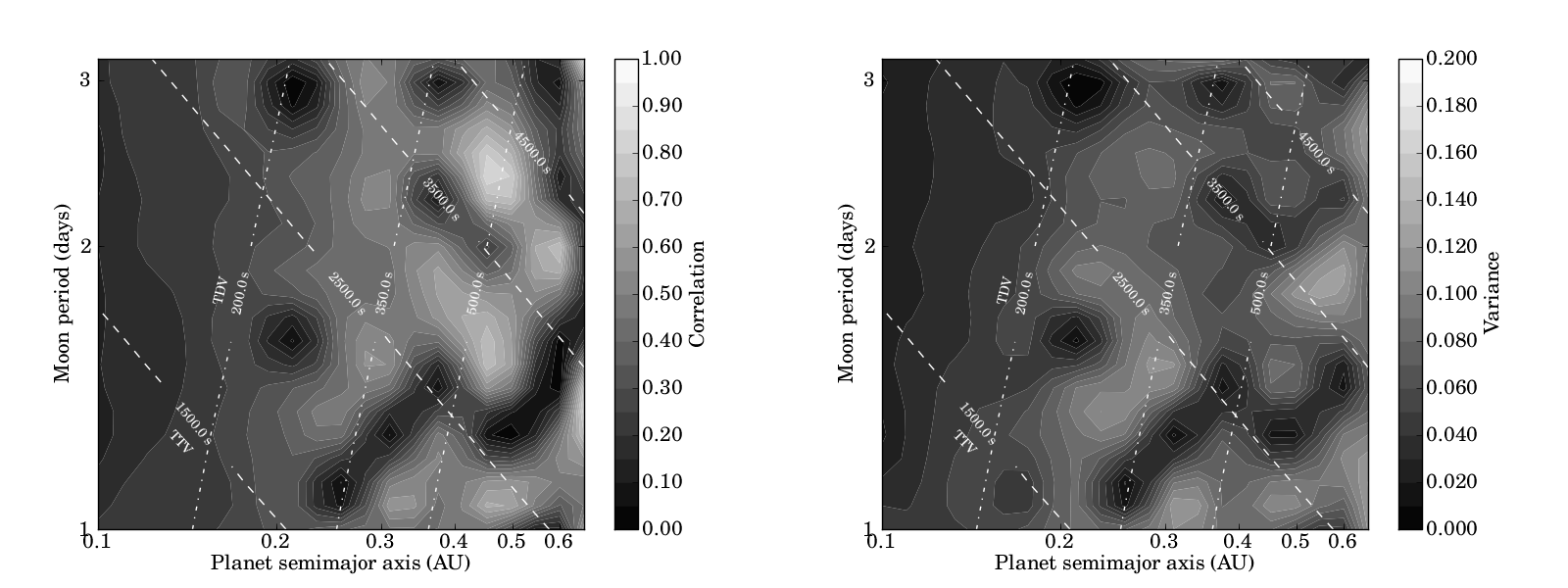} \label{PPPMkp}} \\
(b) Planet semimajor axis versus moon period \\ 
\subfigure{\includegraphics[width=16.5cm]{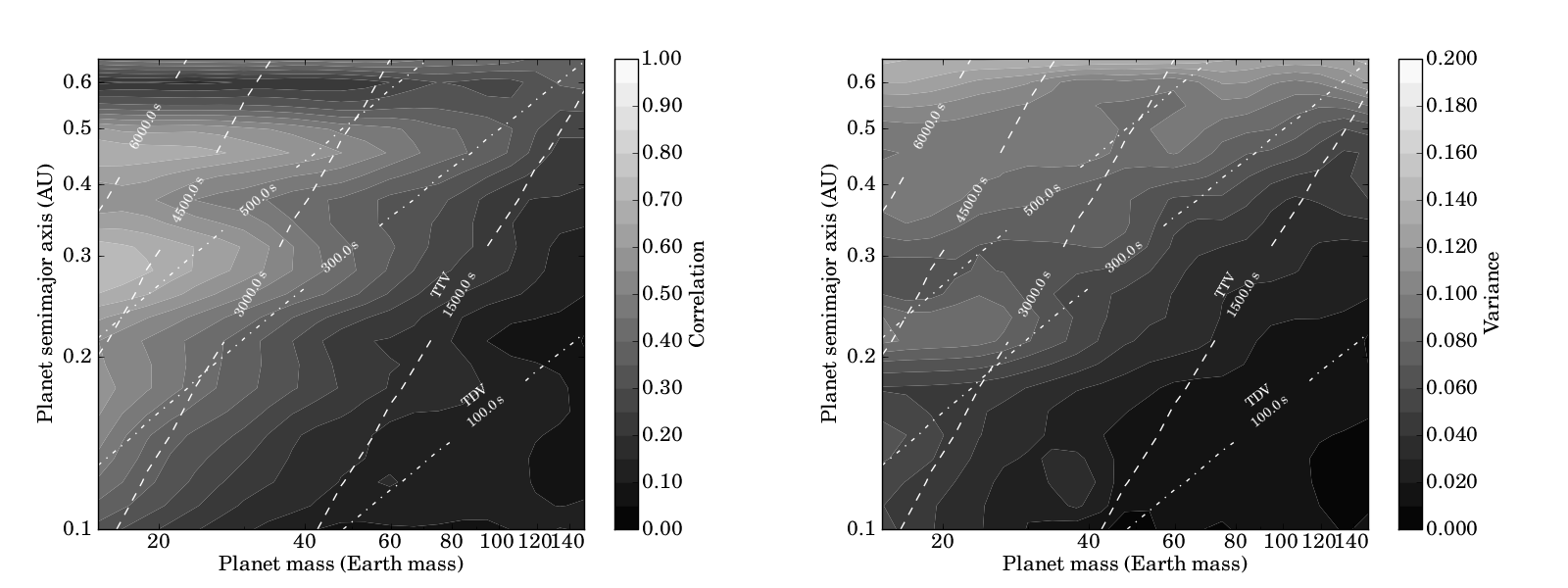} \label{MPPPkp}} \\
(c) Planet mass versus planet semimajor axis \\
\vspace{1cm}
\end{tabular}}
\textbf{Figure 3.} Correlation (left) and variance of correlation (right) between (a) planet mass and moon mass, (b) planet semimajor axis and moon period, (c) planet mass and planet semimajor axis, (d) moon mass and planet semimajor axis (next page), (e) planet mass versus moon period, and (f) moon mass and moon period of the light curves. The contour is averaged over other two variable. The RMS amplitude of the TTV signal (dashed) and RMS amplitude of the TDV signal (dot-dashed) in units of seconds are presented. The cold spots in (b) indicate the data with the planet period in resonance with the moon period (See Section 5.2).
\end{minipage}
\end{figure*}
\begin{figure*}
\begin{minipage}{16.5cm}
{\centering
\begin{tabular}{c}
\subfigure{\includegraphics[width=16.5cm]{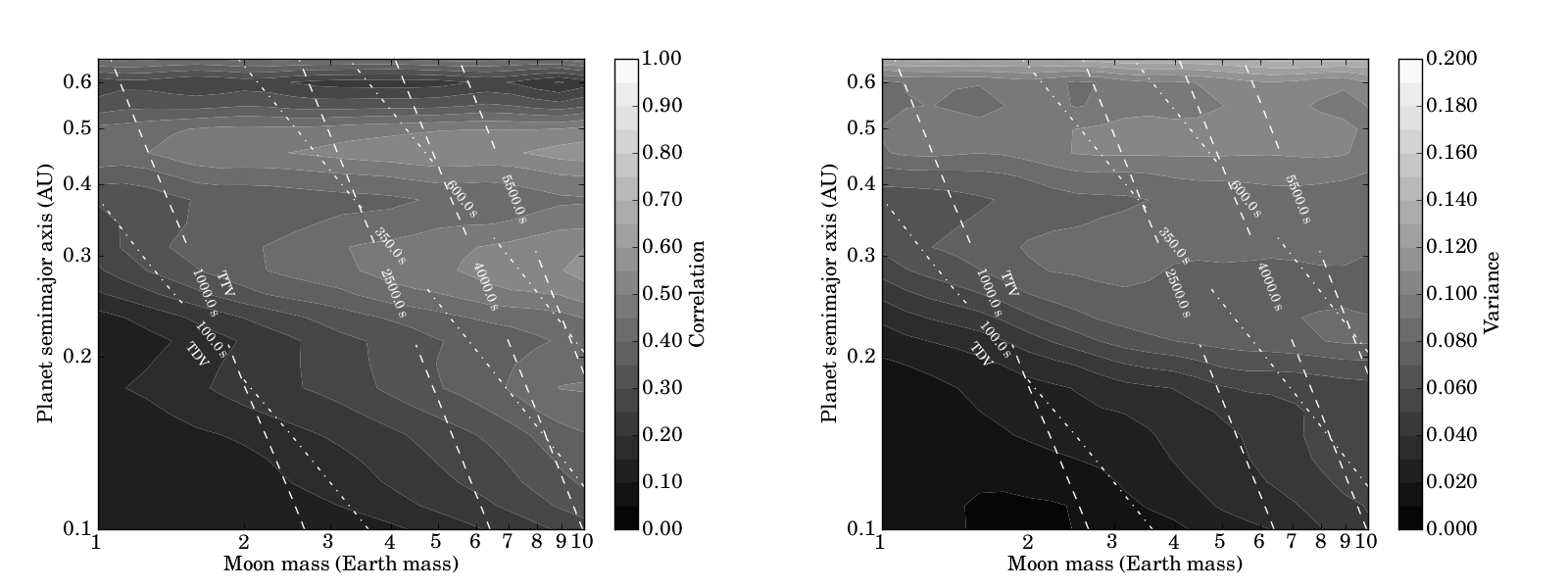} \label{MMPPkp}} \\
(d) Moon mass versus planet semimajor axis \\
\subfigure{\includegraphics[width=16.5cm]{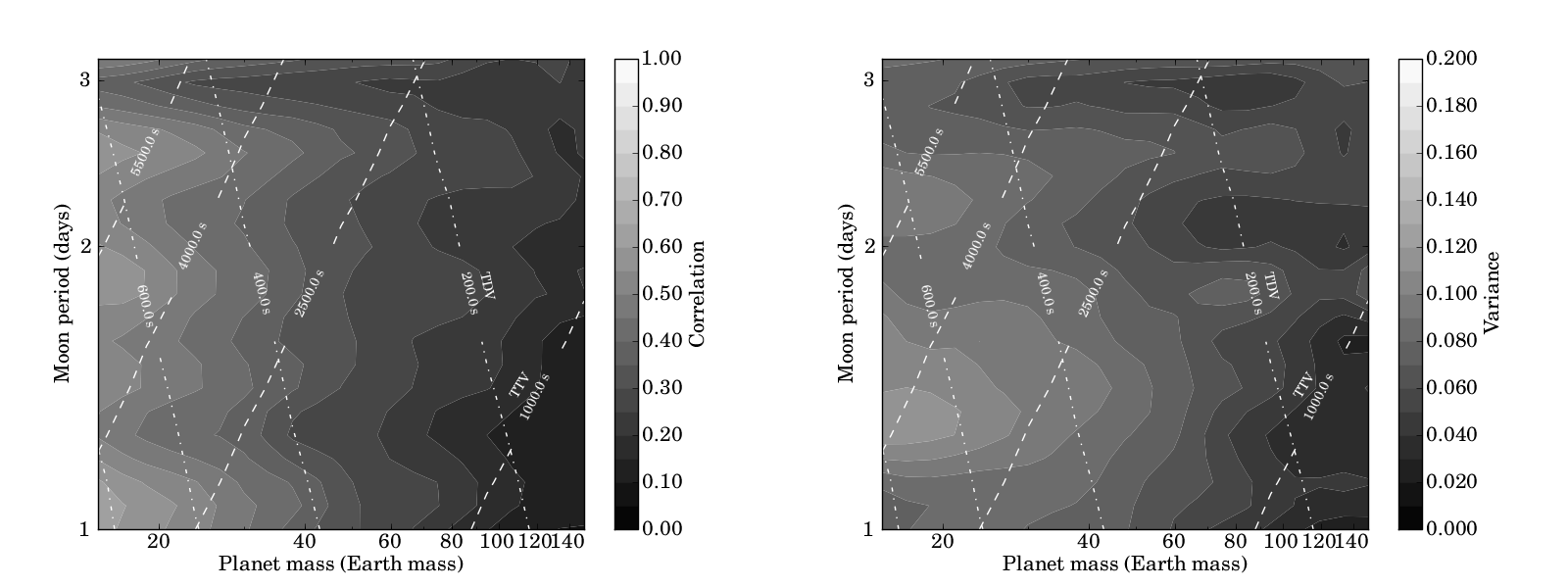} \label{MPPMkp}} \\
(e) Planet mass versus moon period \\
\subfigure{\includegraphics[width=16.5cm]{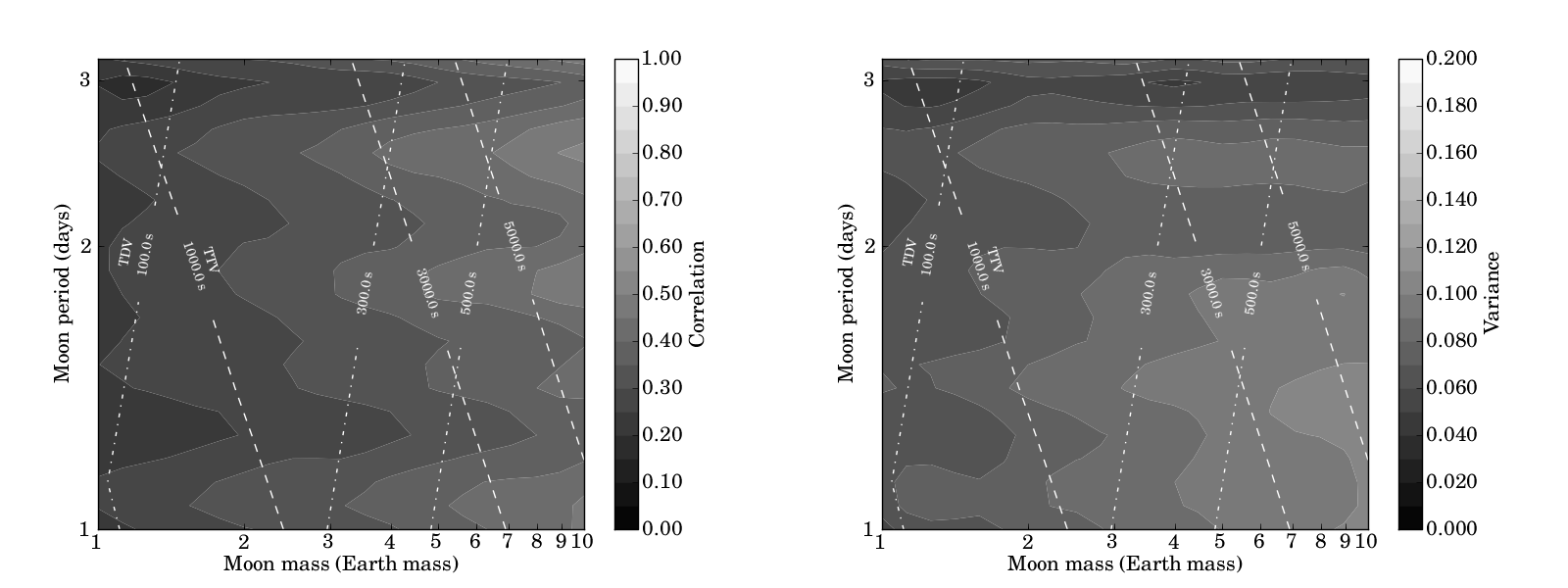} \label{MMPMkp}} \\
(f) Moon mass versus moon period \\
\end{tabular}}
\caption{Continued}
\label{contourKP}
\end{minipage}
\end{figure*}

\begin{landscape}

\begin{figure}
{\centering
\includegraphics[width=1.3\textwidth]{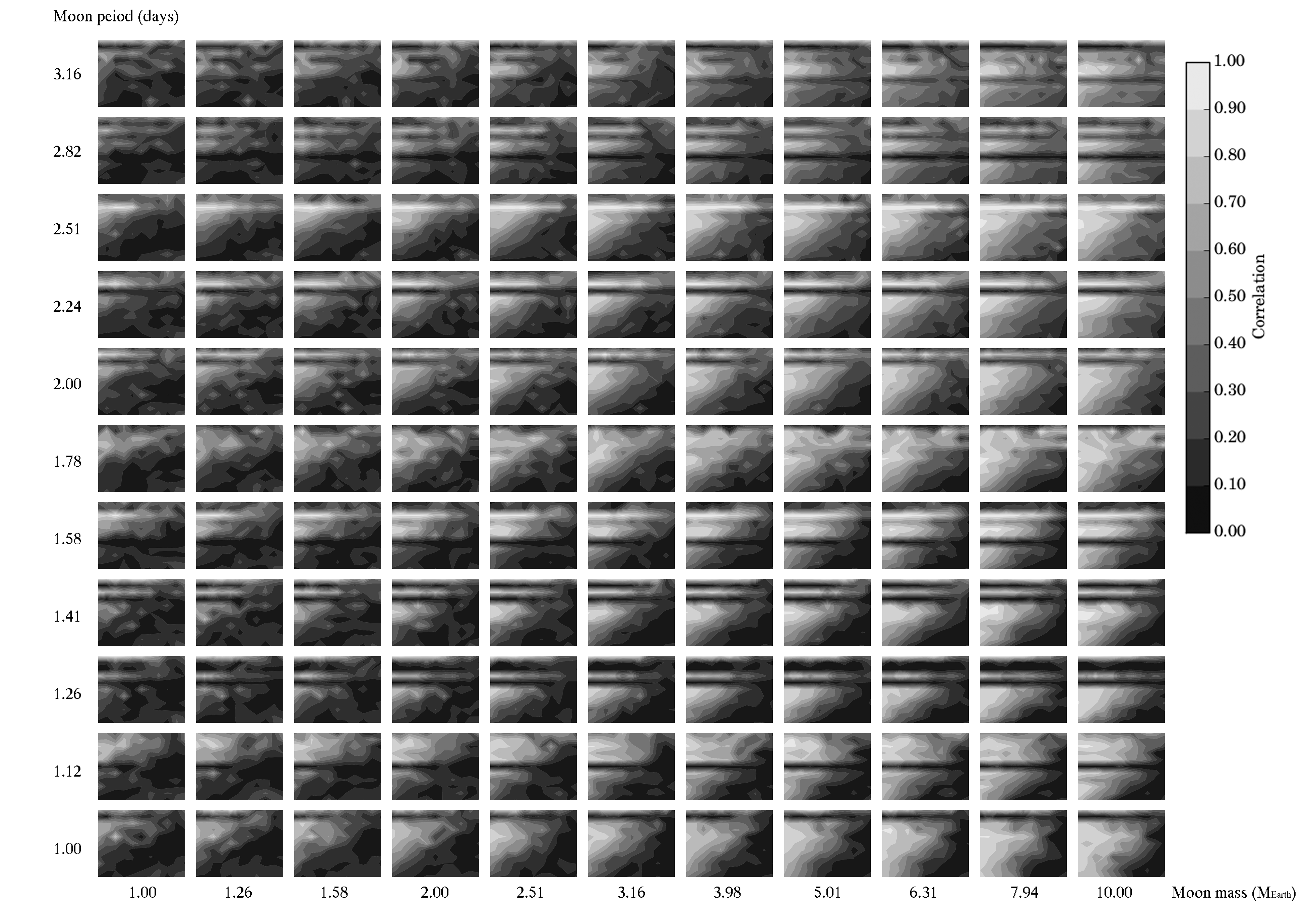}}
\caption{The correlation plots of planet separation versus planet mass on a logarithmically spaced grid. The axis ranges are the same as the range used in Figure~\ref{MPPPkp}.}
\label{MPPP}
\end{figure}

\end{landscape}
\begin{landscape}

\begin{figure}
{\centering
\includegraphics[width=1.3\textwidth]{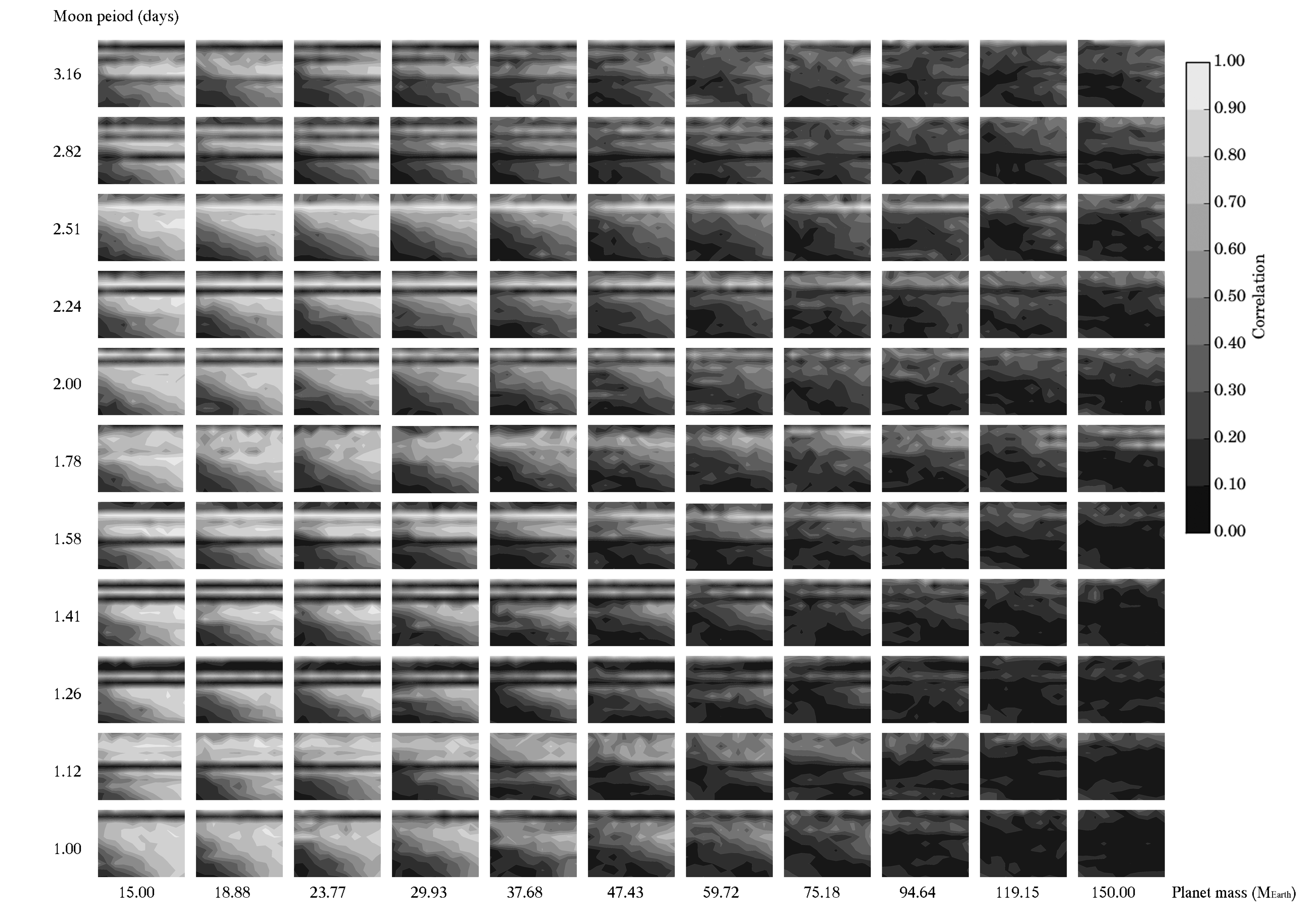}}
\caption{The correlation plots of planet separation versus moon mass on a logarithmically spaced grid. The axis ranges are the same as the range used in Figure~\ref{MMPPkp}.}
\label{MMPP}
\end{figure}

\end{landscape}

\section{Conclusion}

In this work, the light curves of a transiting exoplanet with an exomoon were implemented for the purpose of determining detectability of exomoons. The {\it Kepler} photometric noise was modelled to the light curve in order to simulate the data from {\it Kepler}. Measuring the detectability was done by phase-correlation between TTV and TDV signals. TTV and TDV always exhibit a 90-degree phase shift, therefore, the TTV$^{2}$ signal is linear with the TDV$^{2}$ signal. The Pearson product-moment correlation coefficient was used to determine the detectability of signals. 

3-year {\it Kepler} light curves of 146,410 systems with various configurations were simulated. For each extrasolar planet system, the giant planets and their rocky satellites were placed in the habitable zone of 0.5 $M_{\odot}$ dwarf stars. Their masses and periods were selected logarithmically. For simplicity, edge-on circular orbits with 10 random initial orbital phases were used. From analysing simulated light curves, the detectability of exomoons increases significantly with the moon's mass and decreases with increasing mass of planets. Moreover, the correlation coefficient of systems with the same planet mass or moon mass decreases with the planet semi-major-axis which can be explained by the correlation between the detectability and intensity of TDV signals which are weaker than TTV signals. Exomoon periods in resonance with the planetary orbital period may prevent detection due to the constant observed planet orbital phase. The effects of intrinsic stellar variation (red noise) of M-dwarf reduce the detectability by 0.1. For simulation with red noise of system with planet masses less than around 25 $M_{\oplus}$, 25-50\% of simulated systems with 8-10 $M_{\oplus}$ moon have correlations high enough to confirm the presence of an exomoon.

In conclusion, the detectability of habitable exomoons is ultimately determined by the ability of {\it Kepler} to detect the TDV signal as this is typically weaker than the TTV signal. Resonance between  planet period and moon period can prevent detection of some exomoon configurations. Planets with mass less than around 25 $M_{\oplus}$ and star-planet orbital separation more than 0.3 AU should be the best candidates to detect exomoons in an M-dwarf system. Finally, exomoons in the habitable zone of an M-dwarf system can be detected by 3-years {\it Kepler}'s data and the number of detected extrasolar moons should increase with the number of detected transiting exoplanets in the near future. 

\section*{Acknowledgments}

The authors acknowledge the anonymous referee for his or her valuable suggests that helped to improve the paper. SA thank D. M. Kipping for useful conversation and gratefully acknowledges the support from the Thai Government Scholarship.

\footnotesize
\bibliographystyle{mnras} 
\bibliography{MNRASrefs}

\appendix

\bsp

\label{lastpage}

\end{document}